\documentclass[twocolumn,superscriptaddress]{revtex4-2}
\usepackage{epsfig}
\usepackage{graphicx}
\usepackage{amsmath}
\usepackage{amssymb}
\usepackage{mathrsfs}
\usepackage{verbatim}
\usepackage{dutchcal}
\usepackage{comment}
\usepackage{float}
\usepackage[normalem]{ulem}
\usepackage{xcolor}
\usepackage{slashed}

\newcounter{fig}

\newcommand{\bea}{\begin{eqnarray}}
\newcommand{\eea}{\end{eqnarray}}
\newcommand{\be}{\begin{equation}}
\newcommand{\ee}{\end{equation}}

\def\({\left(}
\def\){\right)}

\newcommand{\re}[1]{(\ref{#1})}

\newcommand{\tr}{\mbox{Tr}}

\def\rlx{\relax\leavevmode}
\def\IR{\rlx\hbox{\rm I\kern-.18em R}}
\def\one{\hbox{{1}\kern-.25em\hbox{l}}}

\newcommand{\eqn}{\begin{eqnarray}}
\newcommand{\eqnx}{\end{eqnarray}}

\begin{document}

\title{Fermion states localized on a self-gravitating
Skyrmion}

\author{
Vladimir Dzhunushaliev }
\affiliation{ Department of Theoretical and Nuclear Physics,
Al-Farabi Kazakh National University, Almaty 050040, Kazakhstan }
\affiliation{ Institute of Nuclear Physics, Almaty 050032,
Kazakhstan } \affiliation{ Academician J.~Jeenbaev Institute of
Physics of the NAS of the Kyrgyz Republic, 265 a, Chui Street,
Bishkek 720071, Kyrgyzstan } \affiliation{ International
Laboratory for Theoretical Cosmology, Tomsk State University of
Control Systems and Radioelectronics (TUSUR), Tomsk 634050, Russia
}
\author{Vladimir Folomeev}
\affiliation{ Institute of Nuclear Physics, Almaty 050032,
Kazakhstan } \affiliation{ Academician J.~Jeenbaev Institute of
Physics of the NAS of the Kyrgyz Republic, 265 a, Chui Street,
Bishkek 720071, Kyrgyzstan } \affiliation{ International
Laboratory for Theoretical Cosmology, Tomsk State University of
Control Systems and Radioelectronics (TUSUR), Tomsk 634050, Russia
}

\author{Jutta Kunz}
\affiliation{Institute of Physics, Carl von Ossietzky University
Oldenburg, Germany Oldenburg D-26111, Germany}

\author{Yakov Shnir}
\affiliation{BLTP, JINR, Dubna 141980, Moscow Region, Russia}
\affiliation{Institute of Physics, Carl von Ossietzky University
Oldenburg, Germany Oldenburg D-26111, Germany}

\begin{abstract}
We investigate self-gravitating solutions of the Einstein-Skyrme
theory coupled to spin-isospin Dirac fermions and consider the
dependence of the spectral flow on the effective gravitational
coupling constant and on the Yukawa coupling. It is shown that the
effects of the backreaction of the fermionic mode may strongly
deform the configuration.
Depending on the choice of parameters,
solutions with positive, negative and  zero ADM mass may arise.
The occurrence of regular anti-gravitating asymptotically flat
solutions with negative ADM mass is caused by the violation of the
energy conditions.
\end{abstract}


\maketitle

\textit{Introduction} -- There are various classical field
theories that admit topologically stable soliton solutions. These
are particle-like regular localized field configurations with
finite energy. Famous examples of such topological solitons in
(3+1)-dimensions are monopoles in the Yang-Mills-Higgs model
\cite{tHooft:1974kcl,Polyakov:1974ek} and
Skyrmions \cite{Skyrme:1961vq,Skyrme:1962vh}
(for reviews, see, for example,
Refs.~\cite{Manton:2004tk,Shnir:2018yzp}).

A peculiar feature of topological solitons is the 
occurrence of fermionic zero modes localized on the soliton.
In particular, there is a link between the
topological charge of the configuration and the number of
zero modes
\cite{Atiyah:1975jf}. 
Fermionic modes localized on solitons were first discussed in the
pioneering work \cite{Caroli:1964}. Later, it was shown that such
bound states exist on kinks
\cite{Jackiw:1975fn,Dashen:1974cj,Chu:2007xh}, domain walls
\cite{Stojkovic:2000ub}, monopoles \cite{Callias:1977cc},
sphalerons \cite{Nohl:1975jg,Boguta:1985ut}, and Skyrmions
\cite{Hiller:1986ry,Kahana:1984be,Balachandran:1998zq}. The
presence of localized fermion modes gives rise to interesting
phenomena like monopole catalysis of proton decay
\cite{Rubakov:1982fp,Callan:1982au},
emergence of superconducting cosmic strings \cite{Witten:1984eb}, 
and appearance of fractional
quantum numbers of solitons~\cite{Jackiw:1975fn,Jackiw:1981ee}.

Here we would like to distinguish between two different types 
of fermionic zero modes that may arise 
in the presence of topological solitons,
depending on their properties.
On the one hand, topological solitons may support exact
fermionic zero modes that are localized on them, independent
of the Yukawa coupling strength, as is the case, for instance, 
for monopoles \cite{Jackiw:1975fn}.
On the other hand, as is the case for Skyrmions, 
such an exact zero mode is not supported by the topological soliton. 
Instead, for Skyrmions there is a spectral flow of the eigenvalues. 
In particular, at some critical value of the Yukawa coupling
a fermionic mode emerges from the positive continuum, crosses zero 
and then slowly approaches the negative continuum as the coupling grows
\cite{Hiller:1986ry,Kahana:1984be,Balachandran:1998zq}.


A major simplification of most of the above studies is that, until
recently, the backreaction of the fermion mode on the soliton was not taken into account. 
This assumption is justified in the weak coupling limit.
However, as the Yukawa coupling increases, the effects of the
backreaction can be significant, see e.g.
Refs.~\cite{Gani:2010pv,Amado:2014waa,Klimashonok:2019iya,Perapechka:2019vqv,Perapechka:2018yux,Perapechka:2019upv,Campos:2022flw,Gani:2022ity,Bazeia:2022yyv,Saadatmand:2022htx,Weigel:2023fxe}.

As compared to boson fields, fermion fields have attracted less
attention in General Relativity. Although solutions of the Dirac
equation in curved spacetime were constructed many decades ago
\cite{Taub:1937zz}, the consideration of self-gravitating fermions
still remains somewhat obscure, because the Dirac field 
can not be treated on a classical level.
Instead, one needs to retain its basic quantum character.
Typically this is done 
by complying with the Pauli exclusion principle.
The Dirac field is then
treated in terms of normalizable quantum wave functions,
and the appropriate occupation number of the relevant 
fermion mode(s) is imposed.

In particular, one then makes use of the approximation that 
(i)~only single-particle fermion states are considered, 
(ii)~second quantization of the fields is ignored, 
and (iii)~gravity is treated purely classically. 
Under these assumptions, for instance, the
fermion level crossing in the background of the
Einstein-Yang-Mills sphaleron was considered in \cite{Volkov:1994tp}. 
It turned out that self-gravitating
spinor fields may give rise to some interesting phenomena in the
cosmology of the accelerating Universe
\cite{Armendariz-Picon:2003wfx,Cai:2008gk}. 
It was shown
that the Einstein-Dirac equations support regular localized
solitonic solutions \cite{Finster:1998ws}, the so-called Dirac stars
\cite{Herdeiro:2019mbz,Dzhunushaliev:2018jhj,Dzhunushaliev:2019kiy,Blazquez-Salcedo:2019uqq,Herdeiro:2021jgc}.
Moreover, it was demonstrated that the backreaction of self-gravitating fermions may
significantly affect the metric and, in particular, allow for (traversable)
wormholes~\cite{Blazquez-Salcedo:2020czn,Bolokhov:2021fil,Konoplya:2021hsm}.

The main objective of the present Letter is to examine subject
to the above stated approximation a similar
system of a spherically symmetric spin-isospin fermionic mode
localized on the self-gravitating Skyrmion, and to study
the spectral flow in this system, where for consistency
the backreaction of the fermions on the Skyrme field and on the
metric is taken into account.

\textit{The Model} -- We consider the (3+1)-dimensional
Einstein-Skyrme system, coupled to a 
Dirac field carrying spin and isospin
\be S = \int d^4 x \sqrt{-\cal g}
\left(-\frac{R}{16 \pi G} + {\cal L}_m \right)\ , \label{Lag-tot}
\ee where the gravitational part is the Einstein-Hilbert action,
$\cal g$ is the determinant of the metric, $R$ is the curvature
scalar, and $G$ is Newton's constant~\footnote{We use natural
units with $c=\hbar=1$ throughout the Letter.}. The Lagrangian of
the matter fields ${\cal L}_m$ is given in terms of the Skyrme
field $U\in SU(2)$ \cite{Skyrme:1961vq,Skyrme:1962vh} minimally
coupled to the Dirac isospinor fermions $\psi$,  
and
${\cal L}_m= {\cal L}_{\text{Sk}}+{\cal L}_{\text{sp}} + {\cal
L}_{\text{int}}$. We consider the usual Lagrangian of the Skyrme
model without a potential term \be {\cal L}_{\text{Sk}}=-
\frac{f_\pi^2}{4}\tr \(\partial_\mu U\,\partial^\mu U^\dagger\)+
\frac{1}{32a_0^2} \tr \(\left[\partial_\mu U U^\dagger,
\partial_\nu UU^\dagger\right]^2\). \label{SkLag} \ee Here $f_\pi$
and $a_0$ are the parameters of the model with dimensions $[f_\pi]
=L^{-1}$ and $[a_0]= L^0$, respectively.

The matrix-valued field $U$ can be decomposed into the scalar
component $\phi_0$ and the pion isotriplet $\phi_n$ via $
U=\phi_0\,\mathbb{I}+i\phi_n \tau_n$, where $\tau_n$ are the Pauli
matrices, and the field components $\phi^a = (\phi_0,\phi_n)$ are
subject to the sigma-model constraint, $\phi^a \cdot \phi^a = 1 $.

The Dirac Lagrangian is
$$
{\cal L}_{\text{sp}} =
 \frac{\imath}{2}\left[
(\hat {\slashed{D}}\bar \psi) \psi - \bar \psi \hat {\slashed D}
\psi \right] - m\bar \psi \psi \, ,
$$
where $m$ is a bare mass of the fermions, $\gamma^\mu$ are the
Dirac matrices in the standard representation in a curved
spacetime, $\hat {\slashed D} = \gamma^\mu \hat D_\mu $, and the
isospinor covariant derivative on a curved spacetime is defined as
(see, e.g., Ref.~\cite{Dolan:2015eua}) $ \hat D_\mu \psi =
(\partial_\mu - \Gamma_\mu )\psi $. Here $\Gamma_\mu$ are the spin
connection matrices~\cite{Eguchi:1980jx,Dolan:2015eua}. 
In the numerical calculations
we restrict ourselves to the case of fermions with zero
bare mass, $m=0$.

Finally, the Skyrmion-fermion chiral interaction Lagrangian is
$$
{\cal L}_{\text{int}}=h \bar \psi \, U^{\gamma_5}\, \psi, \qquad
U^{\gamma_5} \equiv \frac{\mathbb{I} + \tilde \gamma_5}{2}U +
\frac{\mathbb{I} - \tilde \gamma_5}{2}U^\dagger\, ,
$$
where $h$ is the Yukawa coupling constant and $\tilde \gamma^5$ is
the corresponding Dirac matrix in curved spacetime.

It is convenient to introduce the dimensionless radial coordinate
$\tilde r= a_0 f_\pi  r$,  the effective gravitational coupling
$\alpha^2=4\pi G f_\pi^2$, and to rescale the Dirac field, the
Yukawa coupling constant and the bare fermion mass as $\psi \to
\psi/\sqrt{a_0 f_\pi^3 }$,  $h \to h/(a_0 f_\pi)$, and  $m \to
m/(a_0 f_\pi)$, respectively.

To construct spherically symmetric solutions of the model \re{Lag-tot} 
we implement the above set of assumptions.
We treat the gravitational field purely classically and
employ Schwarzschild-like coordinates with a 
spherically symmetric metric, following closely the usual
considerations of self-gravitating Skyrmions (see, e.g., Refs.
\cite{Glendenning:1988qy,Bizon:1992gb,Heusler:1991xx,Heusler:1993ci}),
$
    ds^2 = \sigma^2 (r) N(r) dt^2 - \frac{dr^2}{N(r)} - r^2 \left(
        d \theta^2 + \sin^2 \theta d \varphi^2
    \right)
$.

Likewise, we only consider the scalar fields constituting
the soliton on a purely classical level.
For the static spherically symmetric Skyrmion of topological
degree one, we then make use of the usual hedgehog parametrization $
 U = \cos \left( F(r)\right)  \mathbb{I} + \imath \sin \left( F(r)\right)  \left(
        \sigma^a n^a\right) \, ,
$ where $n^a $ is the unit radial vector. 

The isospin carrying Dirac field is treated by a normalized quantum wave function,
and its fermionic nature is imposed at the level of the occupation number, 
in accordance with Pauli’s exclusion principle.
Thus we consider here a normalized single particle state to describe 
the spectral flow of the system consisting of a Skyrmion with topological number
one, that is interacting with a single isospinor fermion.
The spherically symmetric Ansatz for the isospinor fermion field localized on the
Skyrmion can be written in terms of two $2\times 2$ matrices
$\chi$ and $\eta$~\cite{Jackiw:1975fn,Jackiw:1976xx} as 
$\psi = e^{-\imath\omega t}
\begin{pmatrix}
        \chi \\
        \eta
    \end{pmatrix}
$ with
$$
\chi =
    \frac{u(r)}{\sqrt{2}} \begin{pmatrix}
        0   &   -1 \\
        1   &   0
    \end{pmatrix}, \quad
  \eta =
     \imath \, \frac{v(r)}{\sqrt{2}} \begin{pmatrix}
        \sin \theta e^{- \imath \varphi}    &   - \cos \theta \\
        - \cos \theta    &   -\sin \theta e^{\imath \varphi}
    \end{pmatrix} .
$$
Here $u(r)$ and $v(r)$ are two real functions of the radial
coordinate only, and $\omega$ is the eigenvalue of the Dirac
operator.

\begin{figure}[t]
    \begin{center}
    \includegraphics[width=1.\linewidth]{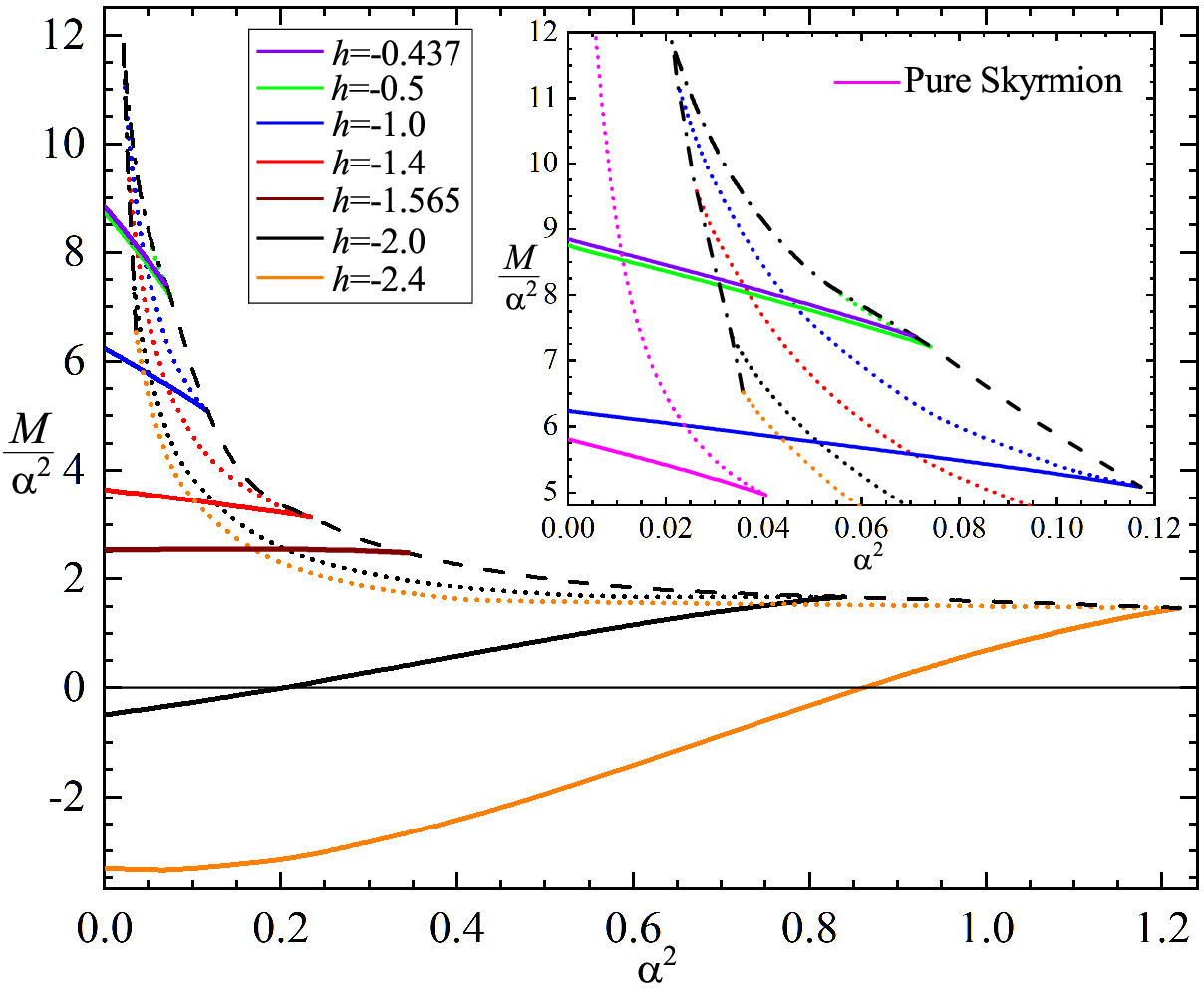}
                \vspace{-1.cm}
    \end{center}
    \caption{
    The ADM mass of the self-gravitating Skyrmion-fermion system is shown as a function of the effective gravitational coupling $\alpha^2$ for several values of $h$.
    The inset shows the behavior of the mass for small $\alpha^2$ in the neighbourhood of the maximum mass obtained.
            }
    \label{M_alpha}
\end{figure}

Varying the total reduced action of the spherically symmetric
self-gravitating Skyrmion coupled to the isospin fermion with
respect to the functions $F,u,v,N,\sigma$, we get the set of five
coupled mixed order ordinary differential equations supplemented
by the normalization condition of the single localized fermion mode $
\int dV\, \psi^\dagger \psi=1$. This system is solved numerically
together with the constraint imposed by the normalization
condition. The boundary conditions are found from the asymptotic
expansion of the solutions on the boundaries of the domain of
integration together with the assumption of regularity and
asymptotic flatness. The Skyrmion profile function $F(r)$
corresponds to the configuration of topological degree one.

\textit{Numerical results} -- In the decoupled limit $(h\to 0)$
the dependence of the regular self-gravitating Skyrmion on the
effective gravitational coupling $\alpha^2= 4 \pi G f_\pi^2$ is
well known. There are two branches of solutions which are
characterized by their limiting behavior as $\alpha$ tends to zero
\cite{Glendenning:1988qy,Bizon:1992gb,Heusler:1991xx,Heusler:1993ci}.
The first branch originates from the flat spacetime Skyrmion (see
the solid purple curve labeled as ``Pure Skyrmion'' in the inset
of Fig.~\ref{M_alpha}). It extends up to a maximal value
$\alpha_{\text{max}}^2 \approx 0.0404$, where it bifurcates with
the second, upper mass branch (the dotted purple curve in the
inset of Fig.~\ref{M_alpha}). The second (backward) branch extends
down  to the limit $\alpha\to 0$ which is approached as $f_\pi \to
0$. Thus the sigma-model term in the Skyrme Lagrangian \re{SkLag}
is vanishing and the configuration tends to the scaled
Bartnik-McKinnon (BM) solution.

\begin{figure}[t]
    \begin{center}
        \includegraphics[width=.97\linewidth]{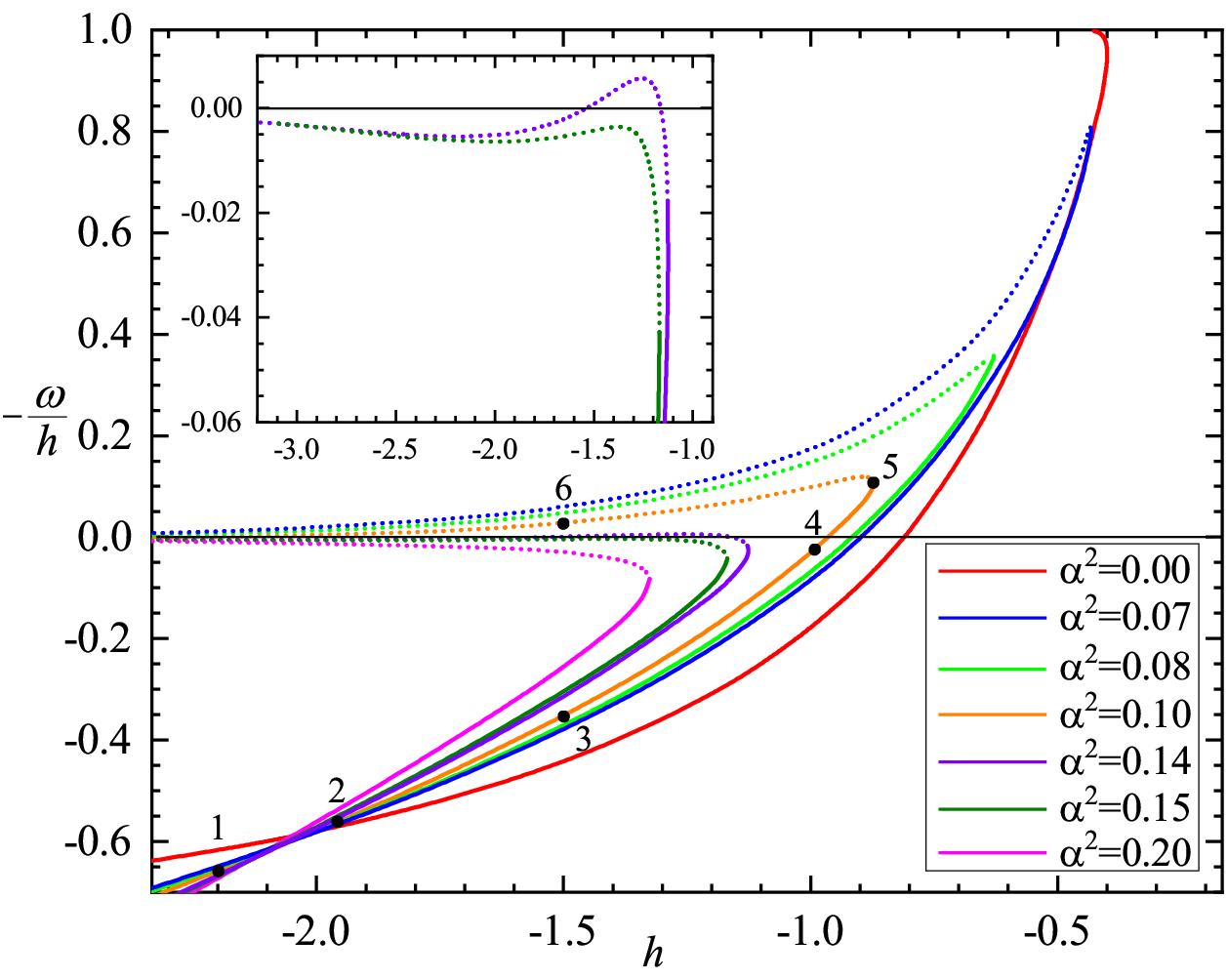}
        \vspace{-.65cm}
    \end{center}
    \caption{
        The normalized eigenvalue of the localized fermionic states is shown as a function of the Skyrmion-fermion
        coupling $h$ for several values of $\alpha^2$.
        The points 1-6 along the curve $\alpha^2=0.1$ correspond to the values of the parameters $\omega$ and $h$
        for which profiles of the matter and metric fields are shown in Fig.~\ref{fig_field_distr}.
        The point 2 corresponds to a configuration with ADM mass $M=0$ (cf.~Fig.~\ref{Mh}). The inset shows the behavior of the curves for $\alpha^2=0.14$ and $0.15$ in the neighbourhood of $\omega=0$.
        }
    \label{fig_freq_coupling}
\end{figure}

In the presence of the fermions, the limit $\alpha=0$ with $G=0$
corresponds to the fermionic mode localized on the Skyrmion in
Minkowski spacetime. This mode emerges from the positive
continuum at some critical value of the Yukawa coupling
$h_{\text{cr}}\approx -0.40$. Further increase of the modulus of
the Yukawa coupling decreases 
the scaled eigenvalue $\omega/|h|$ of
the Dirac operator. For some critical value of the coupling $h$
the curve $\omega (h)$ then crosses zero, as seen in
Fig.~\ref{fig_freq_coupling}. There is a single fermionic level
which monotonically flows from positive 
to negative 
values as the coupling decreases.

More generally, there is a family of solutions depending
continuously on two parameters, the Yukawa coupling constant $h$
and the eigenvalue of the Dirac operator $\omega$, for each
particular value of the effective gravitational coupling $\alpha$.
Since the appearance of a single zero crossing fermionic level is
related to the underlying topology of the Skyrme field, we may
expect that, as the self-gravitating configuration evolves towards the 
topologically trivial 
BM solution, this mode undergoes a
certain transition.

For any non-zero value of the gravitational coupling, the
spherically symmetric fermionic mode localized on the Skyrmion is
no longer linked to the positive 
continuum, as seen in
Fig.~\ref{fig_freq_coupling}. Instead, it arises at some
particular value of the Yukawa coupling $h_{\text{max}}(\alpha) <
h_{\text{cr}}$ with  a scaled  eigenvalue $\omega/|h|$ smaller 
than the threshold value. Physically, this situation reflects the
energy balance of the system of a self-gravitating Skyrmion
interacting with the isospinor fermion:
the added gravitational interaction must be
compensated by the force of the Yukawa interaction. Notably, the
spectral flow of the fermionic Hamiltonian bifurcates at this
point: as $h$ decreases below $h_{\text{max}}$, two branches arise
as displayed in Fig.~\ref{fig_freq_coupling}.

\begin{figure}[t]
    \begin{center}
            \includegraphics[width=.95\linewidth]{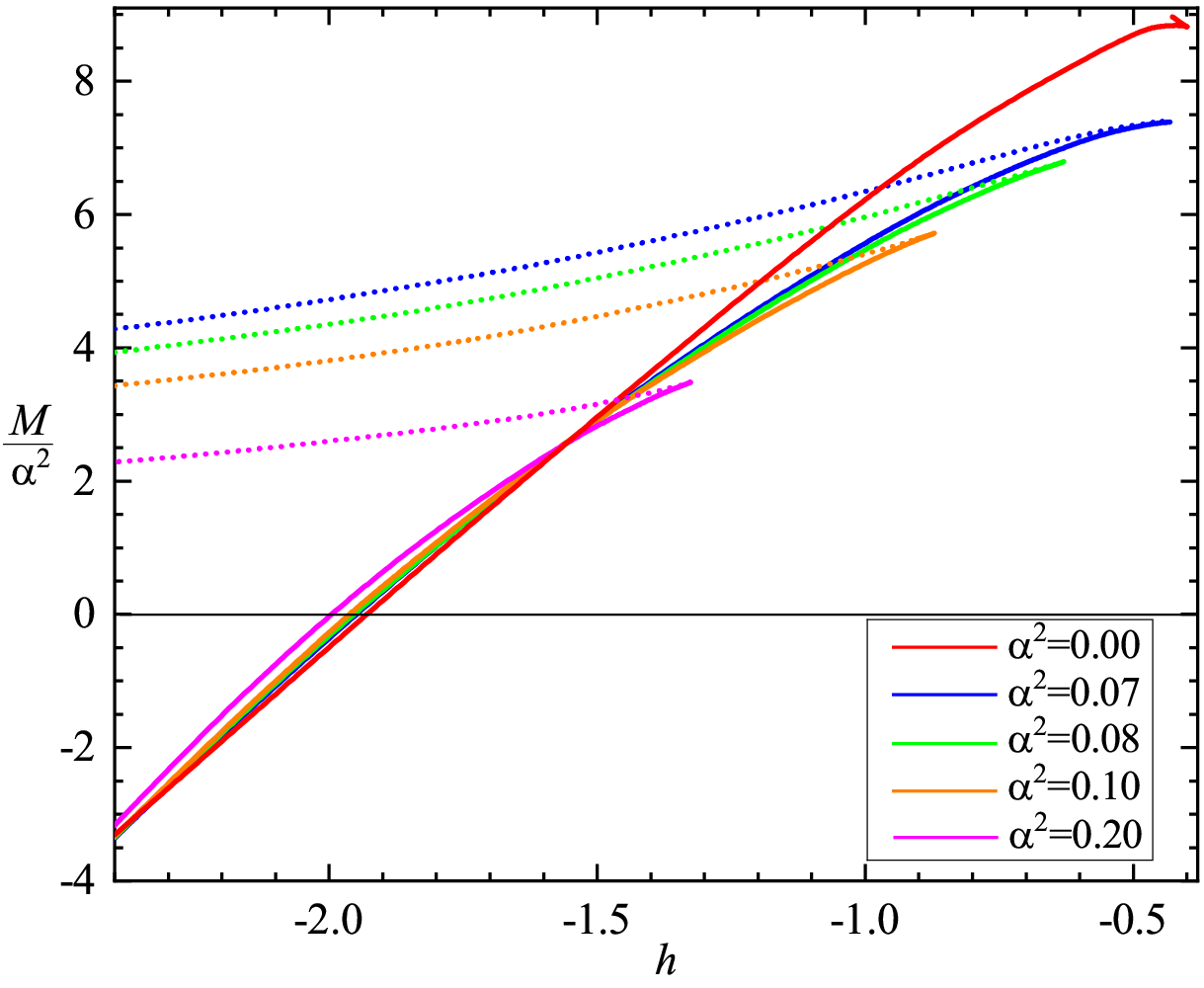}
        \vspace{-.7cm}
    \end{center}
    \caption{
    The ADM mass of the self-gravitating Skyrmion-fermion system is shown
        as a function of the Yukawa coupling $h$ for several values of $\alpha^2$.
        }
    \label{Mh}
\end{figure}

We can understand qualitatively this pattern by analogy with the
appearance of two branches of solutions for self-gravitating
Skyrmions. The evolution of the configuration along one branch is
related to the decrease of the Newton constant $G$, whereas the
second branch may be considered as being obtained by decreasing
the pion decay constant $f_\pi$. In both cases the effective
gravitational coupling $\alpha$ is the same and the configuration
remains in equilibrium for some particular value of the Yukawa
coupling. By analogy with the case of the usual self-gravitating
Skyrmions, we will refer to these branches of the spectral flow as
the ``Skyrmion branch"  and the ``BM branch", respectively. To
distinguish these branches visually, we plot them using solid (for
the Skyrmion branch) and  dotted (for the BM branch) lines in
Figs.~\ref{M_alpha}-\ref{Mh}.

As $\alpha$ remains relatively weak, $\alpha^2\lesssim 0.14$, the
Skyrmion branch of the spectral flow still crosses zero, and the
BM branch slowly approaches zero from
above, as seen in Fig.~\ref{fig_freq_coupling}. Further
increase of the gravitational coupling excludes the zero
eigenvalue of the Dirac operator, and $\omega/|h|$ remains
negative along both branches. On the BM
branch, it then tends to zero from below,
as the Yukawa coupling becomes stronger. Transitional branches
between nodeless branches and those with one node are the branches
for the values of $\alpha\approx 0.14$ when the spectral flow has
two zeros, as seen in the inset of Fig.~\ref{fig_freq_coupling}.

A rather intriguing observation is that the ADM mass of the
coupled configuration becomes \emph{negative} as $h$ decreases
along the Skyrmion branch, as seen in Fig.~\ref{Mh}. On the other
hand, the ADM mass remains always positive along the BM branch.

\begin{figure}[t]
    \begin{center}
        \includegraphics[width=1.\linewidth]{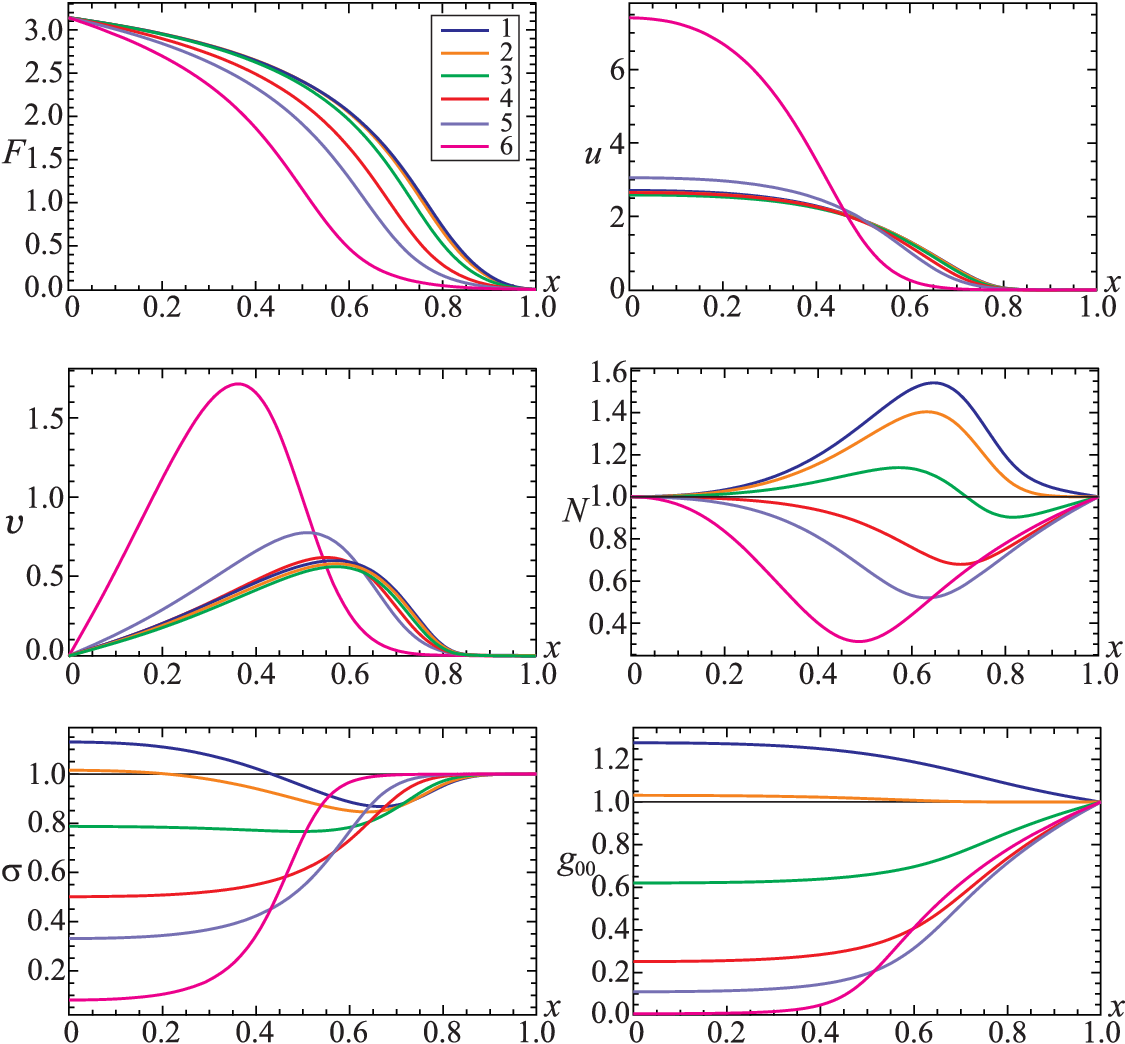}
    \vspace{-.8cm}
    \end{center}
    \caption{The field components of the self-gravitating Skyrmion-fermion system
    are shown as functions of the compactified radial coordinate $ x = \frac{r}{1+r}$
    for $\alpha^2=0.1$ for the points 1-6 in Fig.~\ref{fig_freq_coupling}.
    The solution with negative ADM mass is labeled as 1, while the solution with zero ADM mass is labeled as
    2.
    }
    \label{fig_field_distr}
\end{figure}

In order to get a better understanding, let us now consider the
pattern of evolution of the components of the coupled system.
Fig.~\ref{fig_field_distr} presents the profile functions of the
self-gravitating Skyrmion coupled to the localized fermionic mode
for $\alpha^2=0.1$ and a set of values of the Yukawa coupling.
Note that, as $h$ decreases below the critical value, labeled as
the point 5 in Fig.~\ref{fig_freq_coupling}, the size of the
configuration on the Skyrmion branch increases. As the scaled
eigenvalue $\omega/h$ crosses zero (the point 4) and becomes
positive, the minimal values of the metric functions $N(r)$ and
$\sigma(r)$ are increasing.

A further increase of the strength of the Skyrmion-fermion
coupling for a fixed value of $\alpha$ yields a very unusual
picture: the metric function $N(r)$ increases above unity in the
interior of the Skyrmion, while it becomes less than unity in the
outer region, as seen in Fig.~\ref{fig_field_distr} for the
solution labeled by the point 3 in Fig.~\ref{fig_freq_coupling}.

Notably, the ADM mass of the configuration becomes zero at some
particular value of the Yukawa coupling, as seen in Fig.~\ref{Mh}.
At this critical point the metric component $g_{00}$ is nearly
unity almost everywhere in space and the first derivative of the
metric function $N$ at spatial infinity is vanishing, as displayed
in Fig.~\ref{fig_field_distr}.

As the Yukawa coupling becomes even stronger (the point 1 in
Fig.~\ref{fig_freq_coupling}), the metric function $N(r)$ is
greater than unity, except at the boundaries, and the metric
function $\sigma(r)$ becomes greater than  unity in the inner
region of the configuration. In contrast, the solution on the BM
branch, labeled as the point 6 in Fig.~\ref{fig_freq_coupling},
behaves as expected. The configuration becomes increasingly
localized by the stronger gravitational attraction.

Fig.~\ref{M_alpha} displays the dependence of the ADM mass of the
configurations on the effective gravitational coupling $\alpha^2$.
As before, two branches bifurcate at some maximal value
$\alpha_{\text{max}}$, which increases as the absolute value of
the Yukawa coupling $h$ becomes larger. However, unlike the
self-gravitating Skyrmions in the decoupling limit $h=0$, 
we can not extend
the BM branches of the solutions with a localized fermionic mode 
all the way down to the limit $\alpha\to 0$, as seen in
Fig.~\ref{M_alpha}. For every BM branch, there occurs some limiting
value of $\alpha_{\text{min}}(h)$ for which this branch
terminates in our numerical calculations.
Our 
calculations indicate that there might be 
a critical value of the Yukawa coupling $h\approx -0.79$ for which
the ADM mass attains its maximum value $M/\alpha^2\approx 11.88$.
This would correspond to the minimal value of the gravitational
coupling $\alpha^2_{\text{min}} \approx 0.02$. Correspondingly,
the increase or decrease in the strength of the Yukawa interaction
would increase the minimal value $\alpha_{\text{min}}(h)$, while the
ADM mass of the corresponding limiting systems would decrease, 
as seen in Fig.~\ref{M_alpha}. 
Here the dashed-dotted line then possibly corresponds to
the sequence of limiting configurations on the BM branch which would
terminate at $\alpha_{\text{min}}$ as $h$ varies. In other words,
the presence of the fermions may possibly prohibit the BM limit.
However, further numerical work will be necessary to
confirm or refute such a limit.

On the other hand, the domain of existence of the solutions is
restricted by the bifurcation points, from which the Skyrmion and
BM branches originate. These points are connected by the critical
line (the dashed line in Fig.~\ref{M_alpha}) which restricts the
domain of existence of the solutions from above. This line starts
at $\alpha^2\approx 0.07$ and $h\approx-0.437$ and extends up to
the last calculated point at $\alpha^2\approx 1.22$ and $h=-2.4$.
In turn, in the range $-0.4\lesssim h \lesssim -0.437$, BM branches 
seems to be 
already absent, and 
only Skyrmion branches are found,
which degenerate at $h\approx -0.4$ and $\alpha = 0$ into one flat
spacetime configuration, cf. Fig.~\ref{fig_freq_coupling}.
Curiously, for the fixed value of the Yukawa coupling $h=-1.565$
there is a certain family of configurations on the Skyrmion
branch, whose mass remains approximately constant as $\alpha$
varies up to the bifurcation point (as seen in
Fig.~\ref{M_alpha}).

\begin{figure}[t]
\begin{center}
    \includegraphics[width=1.\linewidth]{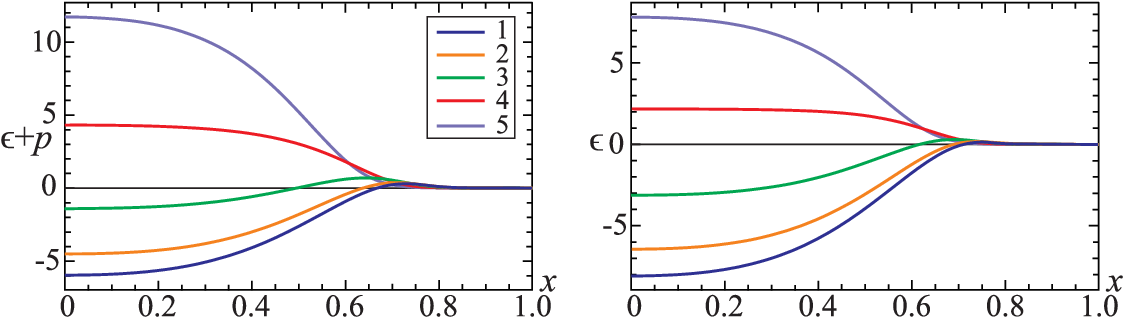}
\vspace{-.8cm}
\end{center}
    \caption{The 
    function $\epsilon + p$
    and the energy density $\epsilon$ for the solutions labeled as 1-5 in Fig.~\ref{fig_freq_coupling}.}
    \label{fig_NEC}
\end{figure}

The unexpected behavior of the solutions of the self-gravitating
Skyrmion-fermion system on the Skyrmion branch is related to the
violation of the null and weak energy conditions, which state that
the stress-energy tensor $T_{\mu\nu}$ satisfies, respectively, the
inequalities $
    T_{\mu\nu} k^\mu k^\nu \geq 0, ~ \text{and} ~ T_{\mu\nu} V^\mu V^\nu \geq 0
$ for any light-like vector $k^\mu$, and for any time-like vector
$V^\mu$ (for a review, see e.g. Ref.~\cite{Rubakov:2014jja}). The
null/weak energy conditions for the self-gravitating coupled
spherically symmetric Skyrmion-fermion system become $
    \epsilon + p \equiv  T^0_0 - T^1_1 \geq 0
$, and the weak energy condition also implies $\epsilon\equiv
T_0^0\geq 0$,
where $\epsilon$ and $p$ are the energy density and radial
pressure, respectively.

In Fig.~\ref{fig_NEC} we display the combination $\epsilon + p$
and the energy density $\epsilon$ for the configurations labeled
by the points $1-5$ on the Skyrmion branch for $\alpha^2=0.1$ in
Fig.~\ref{fig_freq_coupling}. Clearly, the null/weak energy
conditions become violated already for the solution labeled as the
point $3$. Notice also that for all solutions $1-5$ the pressure
$p$ is always positive and the violation of the null/weak energy
conditions is caused only by the negativeness of the energy
density.

\textit{Conclusion } -- Static, spherically symmetric Skyrmions
minimally coupled to gravity are the pioneering example of
solitonic and hairy black hole solutions in General Relativity
\cite{Luckock:1986tr,Glendenning:1988qy,Droz:1991cx,Heusler:1991xx,Bizon:1992gb,Heusler:1993ci}.
The index theorem secures the existence of a bound fermionic mode
localized on the self-gravitating Skyrmion.
However, this mode is not an exact zero mode, independent
of the Skyrmion-fermion coupling. 
Instead the Skyrmion-fermion system exhibits spectral flow.
In this work, we have shown that the localization of the
backreacting fermionic mode may have dramatic consequences, in
particular, the energy conditions may be violated and regular
self-gravitating asymptotically flat solutions with negative and
zero ADM mass may emerge.

\textit{Acknowledgment} -- We are grateful to Ioseph Buchbinder,
Eugen Radu and Alexander Vikman  for inspiring and valuable
discussions. Y.S. would like to thank the
Hanse-Wissenschaftskolleg Delmenhorst for support and
hospitality. J.K. gratefully acknowledges support by DFG project
Ku612/18-1. This research was funded by the Committee of Science
of the Ministry of Science and Higher Education of the Republic of
Kazakhstan (Grant No.~BR21881941).



 \end{document}